\documentclass[11pt]{amsart}
\usepackage[utf8]{inputenc}
\usepackage{graphicx, amsmath, amssymb}
\usepackage[usenames]{color}
\usepackage[normalem]{ulem}
\usepackage[english]{babel}
\usepackage{hyperref}

\begin{document}
\title{Data science and the art of modelling}

\date{May 2018}

\author{
Hykel Hosni         and
        Angelo Vulpiani
}

\address{ (H.H.) Dipartimento di Filosofia, Universit\`a degli Studi
  di Milano\\
and\\
(A.V.) Dipartimento di Fisica, Universit\`a degli Studi di Roma
  Sapienza and\
Centro Linceo  Inderdisciplinare ``Beniamino Segre'', Accademia dei Lincei, Roma
  (Italy).}
\email{hykel.hosni@unimi.it; Angelo.Vulpiani@roma1.infn.it}

\maketitle

\begin{abstract}
  Datacentric enthusiasm is growing strong across a variety of domains. Whilst data science asks unquestionably exciting scientific questions, we argue that its contributions should not be extrapolated from the scientific context in which they originate. In particular we suggest that the simple-minded idea to the effect that data can be seen as a replacement for scientific modelling is not tenable. By recalling some well-known examples from dynamical systems we conclude that data science performs at its best when coupled with the subtle art of modelling.
\end{abstract}

\section{Introduction}
It's been 10 years since Chris Anderson, then the chief editor of the influent technology magazine \emph{Wired}, published an article entitled ``The End of Theory: The Data Deluge Makes the Scientific Method Obsolete'' \cite{and}. Although it does not contain the expression \emph{big data}, which entered common use only a few years later, Anderson's article has quickly become an ideological manifesto of datacentric enthusiasm, articulated along two social and scientific key points.

First: \emph{trust me, it's convenient}. Google has taught us that it is not important to understand why a web page is ``better'' than another -- we just need to pick from the ordering produced by the PageRank algorithm. The convenience of receiving a very simple answer to a potentially very complicated question -- a conscious comparison of an unimaginable quantity of alternatives -- has soon become the key to Google's success. And this probably had a decisive and now consolidated impact in the way we consider rankings and reviews across a variety of contexts, from the choice of a restaurant to the evaluation of the quality of scientific research.

Second: \emph{scientific models are obsolete}. The unprecedented availability of data produced more or less knowingly by all of us, ranging from the great international scientific collaborations, to booking medical appointments online, allows us to rethink radically the relationship between data and the mechanisms generating them. According to Anderson, instead of proceeding by ``conjectures and refutations'' when explaining observations, the data deluge allows us to give up the laborious task of \emph{constructing models} for the phenomena of interest, in favour of the much easier task of analysing the \emph{correlations} identified by sophisticated machine learning algorithms. This  change in perspective is voiced by Anderson (and his followers) with the tone of someone who finally has had the courage to get rid of an old platitude: since ``all models are wrong'', by giving up models we just get rid of obsolete methods, and make room for the full exploitation of cutting edge technologies.

Perhaps it is not surprising that the manifesto of the datacentric enthusiasm, in multiple but analogous forms, has been quickly posted in the offices of many technology start-ups around the world, and has consequently spread throughout various industries. Similarly, it is not surprising that the sirens of simplification and efficiency have quickly seduced a good part of politics and public administration, including public bodies in charge of research policies and funding. What is surprising though,  is the persistence -- in scientific circles -- of the idea that petabytes of data can be self-sufficient,  that all it takes to explain and predict the phenomena of interest in science and applications is the analysis of the correlations in large enough raw data sets. It is surprising because the symmetry of the notion of correlation is an elementary probabilistic fact, as it is the asymmetry of the notion of causation. It is surprising because spurious correlations are pervasive, unless of course you consider as significant a correlation coefficient of 99.26\% between the trend of divorces in Maine and the per capita consumption of margarine in the United States (2000--2009).\footnote{See \url{http://www.tylervigen.com/spurious-correlations} for sources, as well as \cite{cal}.} It is surprising because the very idea of ``raw data'' is at best problematic, since data, or better still, the observations that generate data, are made, not just found. From John Graunt's life table to the monitoring of air quality, the decision of recording, with a given frequency, the values taken by certain random variables inevitably precedes their collection and construction. In this sense, it is very difficult to think about data without them responding to a modelling hypothesis, regardless of the motivation for this hypothesis, be it analogy with other observations, curiosity, or mathematical intuition. What instead seems to animate datacentric enthusiasm, especially in socio-economic areas, is this line of reasoning: \emph{vast amounts of digital traces left behind by billions of people can be obtained and analysed: this ``data'' must be useful to something!} And of course it is, especially in generating revenue for those who sell it.  But in general, in absence of a hypothesis that gives meaning to its collection, those traces are not data. 

It must be noted, and this is our main concern here, that Anderson-like datacentric enthusiasm is not confined to economic institutions\footnote{See for instance the World Economic Forum article ``A brief history of big data everyone should read'', \url{https://www.weforum.org/agenda/2015/02/a-brief-history-of-big-data-everyone-should-read/}. For a critical point of view, see instead the book by Viktor Mayer-Sch\"onberger and Thomas Ramge \cite{may}.}. We find significant traces of it in very prestigious scientific publications. In a work that appeared in \emph{PNAS} in 2015 the group of the ecologist George Sugihara announced an ``Equation-free mechanistic ecosystem forecasting using empirical dynamic modeling'' \cite{ye}. The starting point of this work is the widely shared observation according to which complex natural systems resist the usual mathematical modelling analysis. The proposed solution is, as the title says, the draconian, or better Andersonian: to get rid of equations. This passes a clear and explicit message to the effect that  in describing dynamical systems theories and  models  no longer play the fundamental role they have played for more than three centuries.  The equations that describe the dynamics of the system, in Sugihara and his group's work, are replaced by the method called \emph{empirical dynamical modeling} (EDM), which would make it possible to identify the relevant variables of the ecological model being studied by analysing time series rather than by making scientific hypotheses. Commenting on this article on \emph{Quanta Magazine}, the scientific journalist Gabriel Popkin reports an interview in which Sugihara states that the group's future projects include the application of the ``equation-free'' EDM method to a very wide variety of domains, from finance to neuroscience and genetics \cite{pop}.

But is it really possible to do methodologically sound scientific research starting from ``raw data'', without constructing modelling hypotheses and, therefore, without theory? We think not. Through a brief analysis of the first quantitative model of an ecological system -- the Lotka-Volterra model -- and in particular recalling the way in which it was motivated, constructed and then generalised, the reasons for a deep skepticism with respect to datacentric positions such as Sugihara's will clearly emerge: (i) to construct useful models from data the relevant variables must be chosen accurately and (ii) even when it is theoretically possible to infer the relevant variables from data, the known methods to do so are hindered by severe practical limitations. 
This leaves us with the conclusion that data science performs at its best when it goes hand in hand with the subtle art of constructing models.

\section{The subtle art of constructing models}
Some general considerations on mathematical models and their relationship with the phenomena they intend to describe will provide useful preliminaries. To this end, we borrow from ``A dialogue on the application of mathematics'' written in 1965 by Alfr\'ed R\'enyi (1921--1970) \cite[pp.~29--48]{ren}.

The protagonists are King Hieron and Archimedes whose burning mirrors allowed the Syracusans to sink half of the Roman fleet effortlessly. The dialogue begins with Hieron eager to thank Archimedes for this extraordinary military application of mathematics, an eagerness clearly not reciprocated. Quickly the dialogue moves on to the applications of mathematics to concrete problems, and it is here that Archimedes illustrates to his king the central role of models, and the thoughtfulness necessary to master the art of constructing them.

\begin{quote}
First of all, one can construct many mathematical models for the same practical situation, and one has to choose the most appropriate, that which fits the situation as closely as practical aims require (it can never fit completely). At the same time, it must not be too complicated, but still must be mathematically feasible. These are, of course, conflicting requirements and a delicate balancing of the two is usually necessary. You have to approximate closely the real situation in every respect important for your purposes, but lay aside everything which is of no importance for your actual aims. A model need not to be similar to the modeled reality in every respect, only in those which really count. On the other hand, the same mathematical model can be used to fit quite different practical situations. \dots\ In trying to describe such a complicated situation, even a very rough model may be useful because it gives at least qualitatively correct results, and these may be of even greater practical importance than quantitative results. My experience has taught me that even a crude mathematical model can help us to understand a practical situation better, because in trying to set up a mathematical model we are forced to think over all logical possibilities, to define all notions unambiguously, and to distinguish between important and secondary factors. Even if a mathematical model leads to results which are not in accordance with the facts, it may be useful because the failure of one model can help us find a better one \cite[pp.~38–9]{ren}.
\end{quote}

Hieron proves to be quick in grasping the lesson and comments that as in applied mathematics, so in warfare, defeats are fundamental to the understanding of our mistakes.

Turning from one analogy to another, we can say that giving up models could mean giving up the possibility of identifying errors, and therefore correcting them. In this sense, it is no exaggeration to say that models are inevitable in scientific practice and that methodologically sound science is impossible without a model. Even those very complete and elegant descriptions that we call \emph{theories}, such as classical mechanics or electrodynamics, are, in fact, models.

It is precisely classical physics that offers us the simplest situations to analyse. Here, in fact, there is a fairly clear procedure: once you understand the forces involved, you can write differential equations that may be difficult to solve, but that always allow us to obtain some reliable and useful result, for instance with a qualitative analysis or with a numerical investigation. The task of someone wanting to construct a model for biological phenomena, or for medicine, is decidedly more difficult, not to mention social sciences. In these areas there is nothing like Newton's or Maxwell's laws, and therefore constructing a model can only derive from some often profound intuition, perhaps suggested by analogies or empirical observations.

As R\'enyi's Archimedes says straightaway, there are many models for the same phenomenon, but there are also many different types of models, constructed with various motivations and that have proved useful to understand various problems. Even though in what follows we shall only cover in some detail models of time evolution, such as differential equations, it is useful to have an overview, certainly not exhaustive, of the main types in which models can be classified:

I - Very simplified models that give correct answers to some aspects of the problem considered;\\
II - Models by analogy;\\
III - Large-scale models;\\
IV - Models from data.

An important example of class I is Lorenz's model, which emerges from a brutal simplification of fluid dynamics equations, and which made it possible to understand that irregular motion can be due to chaos, which is also present in low-dimensional systems, as Appendix 1 will show.

In class II we find the model of Lotka-Volterra, which we will cover in next section.

In the models of class III we find the effective equations used in meteorology and engineering, in which only the relevant variables of the problem are considered, and to which Appendix 2 is dedicated.

Class IV contains perhaps the most interesting and certainly difficult problems, in which we have to construct equations from data without having a theoretical frame of reference. These are the kind of intriguing and topical problems exemplified by Sugihara and his group's research project mentioned at the beginning.

Let us start then by comparing this point of view with one of the models of population dynamics of great historical and methodological interest.

\section{The Lotka-Volterra model}
At the beginning of the 20th century, Alfred Lotka (1880--1949) observed experimentally some patterns in certain chemical reactions, but these patterns turned out to be non-periodic. This led him to conjecture that in the field of chemistry and in the wider one of biology it is ``unlikely'' that periodic regularities may occur, in the absence of ``structural causes''. The basis of the conjecture is mathematical: in the solution he suggested for the problem of defining the laws that govern those regularities imaginary exponents appear, and this implies that the parameters with which the evolution of the system is described satisfy suitable geometric properties. In nature, instead, according to Lotka, it is very likely that these parameters are not constrained by any specific structural condition. In light of this, the observation of the special cases of periodic regularity remains to be explained. His 1920 work ``Analytical note on certain rhythmic relations in organic systems'' \cite{lot} answers this question by postulating equations \eqref{eq:LV} that we will discuss in a moment, and showing that under the hypothesis that these describe the population dynamics of a system with two competing biological species (plants and herbivores), the population of the two species could have oscillated periodically.

Vito Volterra (1860--1940) came into contact with the analysis of some data from fishing for the period 1903--1923 performed by his future son-in-law Umberto D'Ancona. In particular, the data concerned the presence of cartilaginous fish in the catch of three Adriatic ports: Trieste, Venice and Fiume (now Rijeka, then in Italian territory). D'Ancona observed a clear growth in the proportion of these species during the First World War, a period of very little fishing. Since cartilaginous fish feed on smaller fish, the conjecture is that the suspension of fishing would favour predators. Unaware of Lotka's work, Volterra reasoned by analogy with the kinetic theory: \emph{the big fish ``collide'' with the small fish and with a certain probability the former eats the latter}, and thus came to identify again the equations \eqref{eq:LV} as a model capable of accounting for D'Ancona's observations.\footnote {This is the motivation given by Volterra for his study \cite{vol}:
\begin{quote}
Dr.~Umberto D'Ancona had repeatedly discussed with me statistics he was collecting about fishing during the war and in the periods before and after it, asking me if it were possible to give a mathematical explanation of the results he was obtaining on the percentages of the various species in these different periods. This request led me to pose the problem as I do in these pages and to solve it by establishing various laws whose statement can be found here.
\end{quote}
}

The equations that govern the dynamics of the two populations are as follows:
\begin {equation}
{dx \over dt} = ax - bxy \, \,, \, \,
{dy \over dt} = -cx + dxy, \, \, \label{eq:LV}
\end{equation}
where $x$ and $y$ are, respectively, the numbers of prey and predator individuals, and the constants $a$, $b$, $c$ and $d$ are positive. The linear terms do not need many explanations: assuming unlimited food resources, the absence of predators leads to an exponential proliferation of the prey; analogously, in the absence of prey, predators become extinct. In the fish-related case that aroused Volterra's interest, the non-linear terms have instead the following interpretation: when both cartilaginous fish and their prey increase, the population transfer from prey to predators also increases.

The Lotka-Volterra equations allow us to identify two equilibrium states between populations. The first is the one in which both populations are extinct, and therefore $x = y = 0$. The second one is that in which the two populations coexist:
$$x = c/d,\,\,\, y = a/b.$$

Despite the simplicity of the model, for which in a sense Volterra himself apologises in the introduction to his 1927 work, it is possible to derive a theoretical prediction that is anything but trivial, and in particular the \emph{mathematical explanation} of zoological observations. Suppose the two fish populations are not in equilibrium. Then the abundance of small fish leads to an increase in the population of cartilaginous fish. These, in feeding, will quickly cause a decrease in the population of the other species, until some cartilaginous fish will starve, allowing the repopulation of small fish. And so on, periodically.

Perhaps some readers will have noticed in equation \eqref{eq:LV} the (bilinear) structure similar to that of Boltzmann equation. This is not a coincidence: the nonlinear terms were introduced by Volterra noting the analogy between the prey/predator interaction and the impact of two atoms in the kinetic theory while, as we have seen above, Lotka had chemical reactions in mind -- in fact, it is the same mechanism.

Once the model based on the equations \eqref{eq:LV} has been constructed, we can reason about it in a purely mathematical way, asking ourselves if it is possible to extend the expressivity and therefore the predictive ability of the model itself. For instance, it is natural to wonder if the model only ``works'' with algebraic nonlinearities, or if we can consider the case with $N$ different species $x_1$, \dots, $x_N$. It is worth mentioning that in ecology there is nothing similar to Newton's mechanics, and therefore such generalisations cannot be sought by relying on first principles. However, the analogy with the Lotka-Volterra model and the consistency with the relevant ecological facts emerge as very natural constraints. Thus, Andrey N. Kolmogorov (1903--1987) introduced a generalisation of the Lotka-Volterra equations for the $N = 2$ case of the form
\begin{equation}
{dx \over dt} = x F(x, y) \, \,, \, \, \,
{dy \over dt} = y G(x, y), \label{eq:2}
\end{equation}
Smale considered the case with $N \ge 3$ for a class of possible equations
\begin{equation}
{dx_n \over dt} = x_n M_n(x_1, \dots, x_N). \label{eq:3}
\end{equation}
For the properties of the functions $F$, $G$ and $\{M_n\}$ we cannot appeal to some general theory; however, if we want the model to have only positive solutions (if they are at the initial time), resources to be limited and other conditions that reasonably exist in the ecological sphere to hold, then there are constraints that restrict the class of acceptable models.

Limiting ourselves to nonlinearities of algebraic type, the obvious generalisation of the Lotka-Volterra model is
\begin{equation}
{dx_n \over dt} = a_n x_n (1 - \sum_{j = 1}^N b_{n, j} x_j), \label{eq:4}
\end{equation}
where $a_n$ is positive for prey (herbivores) and negative for predators; the diagonal terms describe the competition between individuals of the same species, the non-diagonal ones specify the type of interactions between the various species, for instance of parasitic, symbiotic or prey/predator type.

Summarising: data, or rather observations, were fundamental for both Lotka and Volterra, but the brief accounts of their motivations and their arguments for choosing equations \eqref{eq:LV} suggest a predominantly heuristic role. Analogy, mathematical intuition and deduction are the main components in the justification of the equations of the model, which both make an effort to show to be \emph{consistent} with fundamental ecological facts. This provides the theoretical basis to derive, this time from the model and therefore mathematically, new conjectures to be submitted to experimental observation. Lotka, for instance, ended the aforementioned work by showing analytically how his model predicts that the amplitude of the oscillation{} between the two species cannot, in the absence of other factors, determine the extinction of both prey and predator. At the same time, we have briefly mentioned some generalisations of the equations \eqref{eq:LV}, which illustrate better than anything the essence of the last piece of advice given by R\'enyi's Archimedes to Hieron: every model is first and foremost a starting point for better models.

\section{How do we construct a model?}
Unfortunately, however, not all of us have the abstraction skills or the mathematical intuition of Lotka, Volterra or Kolmogorov. Thus, it would be desirable to identify a general method that allows us to determine the equations that govern the phenomenon we are interested in. In this sense, the search for algorithmic methods to generate ``theories'' starting from data is perfectly understandable, if only it were possible.

The generalization of Lotka-Volterra model based on the \eqref{eq:4} is relatively simple, since we know (or rather, it is not unreasonable to assume that we know) the ``right variables'' ${\bf X} = (x_1, x_2, \dots, x_N)$ given by the ecological problem we want to tackle, as well as the structure of the equations, taken in analogy with the original model. What remains to be solved, then, is ``only'' the problem of determining the coefficients $\{a_n\}$ and $\{b_{i,j}\}$, something that is not trivial but that does not seem impossible to do, in particular when good experimental data is available.

In general, that is, if we do not have a reference model from which to start, and we do not have a theoretical hypothesis that constraints the choice of the parameters of the model we want to construct, things are far more difficult. In a first approximation, we can distinguish two situations, in increasing order of difficulty:
\begin{enumerate}
\item we know the ``right variables'' ${\bf X}$, but not the structure of the equations, and we only have one (long) time series $\{{\bf X}_1, {\bf X}_2, \dots, {\bf X}_T\}$ available (for the sake of simplicity, we limit ourselves to the discrete time case);
\item we do not know the ``right variables'' and only have time series of some variable $\{u_1, u_2, \dots, u_T\}$.
\end{enumerate}

Let us start by discussing the simplest case, in which we know the right variables ${\bf X}$ and we also have access to the past, that is, we have a time series
$$
\{{\bf X}_1, {\bf X}_2, \dots, {\bf X}_T\}.
$$
A natural way to infer the future ${\bf X}_{T+n}$ with $n = 1,2, \dots$ is to look for an \emph{analogue} in the past, that is, a ${\bf X}_t$ such that $|{\bf X}_t-{\bf X}_T| < \epsilon$ (where $\epsilon$ depends on the desired accuracy). In this case we can ``predict the future'' using the obvious recipe ${\bf X}_{T+n} \simeq {\bf X}_{t+n}$. Of course, if the system is chaotic, the uncertainty about ${\bf X}_{T+n}$ increases exponentially with $n$.

If we can find the desired analogues, then it is possible with some optimisation procedure to construct a model of the form
$${\bf X}_{t+1} = {\bf G}({\bf X}_t).$$
Now, since the existence of an analogue is assured by the Poincar\'e recurrence theorem, it would seem that for the deterministic models of type 1 it is possible to identify a method -- the one just outlined -- to construct models enabling us to infer the future state of a system starting from a sufficiently long time series. However, appearance is deceiving, since the method of analogy has considerable practical limits. It follows in fact from the Kac lemma \cite{kac} that, in order to find an analogue within a precision $\epsilon$, the time series must have length at least $O(\epsilon^{-D})$, where $D$ is the dimension of the system. To be clear: if we want a precision of $1\%$, then the length of the series must be at least $O(10^{2D})$. Clearly, this method allows us to actually  make predictions only if $D$ is not too large, as evidenced by the case of Edward Lorenz (1917--2008) \cite{loratm}, who in the 1960s tried to apply the analogue method to forecast the weather in North America. His conclusion was:
\begin{quote}
In practice, this procedure may be expected to fail, because of the high probability that no truly good analogues will be found within the recorded history of the atmosphere \cite[p.~347]{lor69}.
\end{quote}
The (practical) impossibility of finding an analogue in a series of a few decades is obviously due to the huge value of $D$ in the problem considered by Lorenz.

Let us now move on to the second situation, the most complicated, even though it is very common, that in which we do not know the right variables ${\bf X}$, but we only have a time series of some observable
$$
\{u_1, u_2, \dots, u_T\}.
$$
We are facing the problem of the \emph{phase-space reconstruction}, that is, of determining the variables that describe the phenomenon in which we are interested. The problem was successfully addressed by the Dutch mathematician Floris Takens (1940--2010). He identified the conditions under which, for systems of dimension $D$, the variables obtained with the method known as {\it embedding},
$$
{\bf Y}^{(m)}_t = (u_t, u_{t-1}, \dots, u_{t-m+1})
$$
for $m > 2[D]+1$, can provide a complete description of the system. $D$ is not known a priori and the embedding procedure must be performed by trial and error, that is, by increasing $m$ and hoping for a convergence; unfortunately, if $D$ is too large, it is practically impossible to have a convergence.

In other words, Takens's results identify the conditions below which the situations we have labeled as type 2 can be reduced to those of type 1. This obviously does not allow us to circumvent the practical limitations due to the Kac lemma: the method can only work in low dimension.

It is interesting to recall{} the approach used today to formulate weather forecasts, which are very accurate up to a few days. The method was proposed in the 1920s by Lewis Fry Richardson (1881--1953) -- who had understood that the approach in terms of analogues had no hope of success -- and is essentially the following: the atmosphere evolves in accordance with the equations of hydrodynamics and thermodynamics, so from the present state of the atmosphere, solving (obviously numerically) a system of partial differential equations, it is possible to make a weather forecast. Richardson's visionary project was carried out starting only in the 1950s, with the development of three completely non-trivial ingredients:
\begin{itemize}
\item[a)] the development of effective equations;
\item[b)] fast numerical algorithms;
\item[c)] computers for numerical calculations.
\end{itemize}

Point a) is obviously, from a conceptual viewpoint, the most important aspect. Necessary for meteorology was the fundamental contribution of Charney and von Neumann, who understood that the equations originally proposed by Richardson, though correct, were not suitable for forecasts. The apparently paradoxical reason is that they were \emph{too accurate}, so much so that they also described high frequency waves that are irrelevant in the meteorological field. Richardson's vision, therefore, became feasible not only as a result of technological progresses in the field of computers, but also following the selection of the variables that are relevant to the construction of effective equations, in which, for instance, fast variables do not appear (See Appendix 2).

If we try to generalise and systematise Richardson's intuition, we face predictable and probably insurmountable problems. It is difficult to think of a general method for choosing the ``right'' variables. This is an aspect that is too often overlooked, even though it is certainly one of the most delicate problems. For instance, Onsager and Machlup are explicit in raising the
question:
\begin{quote}
How do you know you have enough variables, for [the system] to be Markovian? \cite[p.~1509]{ons}.
\end{quote}

Similarly, Shang-Keng Ma expresses a caveat of central importance:
\begin{quote}
The hidden worry of thermodynamics is: we do not know how many coordinates or forces are necessary to completely specify an equilibrium state \cite[p.~29]{ma}.
\end{quote}

There are no automatic protocols for this choice, and typically to obtain good results it is necessary to possess the expertise of which Archimedes spoke to Hieron. Further, it is even more difficult to assume that there are protocols capable of generating this choice from petabytes of ``raw data''. Time series, indeed, are such to the extent that one decides to observe them. This brings us back to where we started.

\section{So, can we ``throw away equations''?}
The preceding discussion allows us to justify a deep skepticism about the position of those who claim that we are facing a new scientific revolution, the datacentric one. The possibility of extracting knowledge through the algorithmic analysis of large amounts of data would have, according to this position, created a fourth paradigm, a new scientific methodology to be added to the three already existing: the experimental method, the mathematical approach and the computational one of numerical simulations.

There are even some who, like the computer science guru Chris Anderson we have recalled at the beginning, have gone as far as to claim that
\begin{quote}
faced with massive data, this approach to science -- hypothesize, model, test -- is becoming obsolete \dots\ Petabytes allow us to say: ``Correlation is enough.'' We can stop looking for models \cite{and}.
\end{quote}
As we have mentioned, it is a short step from these hyperbolic slogans to the widespread diffusion of the idea that it is no longer necessary to study general theories, but that it is sufficient to collect data from the Internet of Things, as they call it, cook them with our computer, obviously with software downloaded from the 'net, and thus get everything we need. This idea is widespread but, as we have argued, unsustainable in its deepest aspects.

We have seen that analysing specific problems (such as weather forecasting) with a suitable critical spirit allows us to understand that the recent collection and algorithmic analysis techniques of a great variety and quantity of data, while constituting an interesting scientific challenge, of potential high impact at all levels of society, are not a panacea.

It is necessary to guard against easy enthusiasm, and especially against the temptation of thinking that the solution to all problems, from large scientific projects to medical diagnosis, depends on the development of this interesting and sophisticated technology.

For further reading about ``big data'', see \cite{hey}, \cite{hos}, \cite{lic}; about the history and pioneers of models and their theory, see \cite{bac}, \cite{dah}, \cite{gia}, \cite{gue}, \cite{guepao}, \cite{vul}; about the feasibility of predictions and forecasting, see \cite{cec12}, \cite{cec13}, \cite{lyn}, \cite{wei}.

\section*{Appendix 1. Lorenz's model}
A rather common problem in many applications is the following: given a nonlinear partial differential equation
\begin{equation}
\partial_t \psi({\bf x},t) = {\mathcal L}[\psi({\bf x},t), \nabla\psi({\bf x},t), \Delta\psi({\bf x},t)]   \label{B1}
\end{equation}
where $\psi$ is a vector field, we want to find a set of differential equations that approximate (\ref{B1}). As an example, we may consider the Navier-Stokes equations with $\psi = ({\bf u}, \rho, p, T)$,
where ${\bf u}$, $\rho$, $p$ and $T$ denote, respectively, the velocity field, density, pressure and temperature.

A widely used procedure (the so-called Galerkin method) consists in approximating $\psi({\bf x},t)$ in the form
\begin{equation}
\psi({\bf x},t)=\sum_{n<N} a_n(t) \phi_n({\bf x}),
\label{B2}
\end{equation}
where $\{\phi_n\}$ are suitable orthonormal, complete functions. Substituting (\ref{B2}) in (\ref{B1}), we obtain a set of differential equations for $\{a_n\} $:
\begin{equation}
{d a_n \over dt}=F_n(a_1,a_2,.., a_N)\,\,\, , \,\, n=1,2,...,N.
\label{B3}
\end{equation}

Of course, if we want a good quantitative agreement, then $N$ has to be very large, so that the set of differential equations (\ref{B3}) is a good approximation of (\ref{B1}). This is what is done in meteorology or engineering, where the value of $N$ easily reaches $10^9$ and even more.

In his famous 1963 paper \cite{lor63}, Lorenz, studying the problem of convection in a fluid heated from below, used for $N$ the smallest value that could give non-periodic behaviours, that is $N = 3$: in particular, 2 harmonics for speed and 1 for temperature. Here is his famous model, apparently innocuous:
\begin{equation}
{dx \over dt}= -\sigma x + \sigma y, \,\,\,
{dy \over dt}= -xz+rx-y, \,\,\, 
{dz \over dt}= xy -bz,
\label{B4}
\end{equation}
where $(x, y, z)$ are proportional to $(a_1, a_2, a_3)$, and $\sigma$, $b$ and $r$ are constants related to the properties of the fluid; in particular,
$r$ is proportional to the Rayleigh number.

It is important to emphasise the fact that these equations were not invented, but were obtained, even if with a very brutal truncation, from the equations of fluid dynamics. Of course, the value of $N = 3$ does not allow a quantitative agreement with the original equations.

The importance of Lorenz's model lies in having shown that it is possible to obtain a chaotic behaviour even in low-dimension systems: the complexity of the temporal evolution that occurs in turbulent fluids is not necessarily a mere superposition of many elementary events (say, many Fourier harmonics), but comes from the nonlinear structure of the equations.

\section*{Appendix 2. The role of theory and of right variables in weather forecasting}
In the 1950s, Charney and von Neumann, in the context of the Meteorological Project at the Institute for Advanced Study in Princeton, noticed that the equations originally proposed by Richardson, even though correct, are not suitable for weather forecasting. The apparently paradoxical reason is that they are too accurate, as mentioned above. It was thus necessary to construct effective equations that eliminated the fast variables. The introduction of the filtering procedure, which separates the meteorologically relevant part from the irrelevant one, has a clear practical advantage: numerical instabilities are less severe and therefore a relatively large $\Delta t$ integration step can be used, which allows more efficient numerical calculations.

Besides the computational aspect, it is important to note that with effective equations for the slow dynamics it is possible to identify the most important ingredients, which instead remain hidden in the detailed description in the system given by the original equations. The equations used are called quasi-geostrophic; the simplest case is the barotropic one, in which the pressure depends only on the horizontal coordinates.

To give an idea of the construction of effective equations for slow variables, consider a (rather academic) case in which the status of the system ${\bf X}$ consists of slow variables ${\bf X}_S$, with a characteristic time $O(1)$, and fast variables ${\bf X}_f $ with a characteristic time $O(\epsilon) \ll 1$, which evolve with a set of differential equations
$$
{d{\bf X}_s \over dt} = {\bf F}({\bf X}_s, {\bf X}_f) \,, \, \, \,
{d{\bf X}_f \over dt} = {1 \over \epsilon} {\bf G}({\bf X}_s, {\bf X}_f).
\eqno(A1)
$$
Note that even a numerical study of this problem is not simple: it would require the use of an extremely small $\Delta t$ integration step, that is, much smaller than $\epsilon$.

On the other hand, if we are only interested in slow variables, it is sufficient to write an equation for the ${\bf X}_s$:
$$
{d{\bf X}_s \over dt} = {\bf F}_{\it eff}({\bf X}_s),
$$
which takes into account the effect of fast variables on slow ones; this way, we could use a $\Delta t$ that is not too small.

Unfortunately, there are no systematic procedures to find effective equations, and not even the separation of the variables into slow and fast ones is easy.

Finally, in the case studied by Charney and von Neumann, things are even more difficult since we have partial differential equations.

Perhaps the most famous example of eliminating fast variables is given by the Langevin equation, which describes the motion of a colloidal particle in a liquid. These particles are much larger than the liquid molecules, their dimensions being in the order of microns, and much slower; the effect of the molecules translates into a friction force and a fluctuating force (white noise).

\end{document}